\documentclass[summary]{ursi}

\title{Near Field Optimization Algorithm for Reconfigurable Intelligent Surface}

\author{E. Colella\affref{ref1}\affref{ref2}, L. Bastianelli \affref{ref1}\affref{ref2}, F. Moglie \affref{ref1}\affref{ref2} and V. Mariani Primiani\affref{ref1}\affref{ref2}}

\affiliation{%
  \aff{ref1}{Università Politecnica delle Marche}
  \aff{ref2}{Consorzio Nazionale Interuniversitario delle Telecomunicazioni (CNIT)}
}

\begin{document}

\maketitle

\begin{abstract}
Reconfigurable intelligent surface (RIS) is a type of wireless communication technology that uses a reconfigurable surface, such as a wall or building that is able to adjust its properties by an integrated optimization algorithm in order to optimize the signal propagation for a given communication scenario. As a reconfiguration algorithm the multidimensional optimization of the GNU scientific library was analyzed to evaluate the performance of the smart surface in the quality of signal reception. This analysis took place by means of electrodynamic simulations based on the finite difference time domain method. Through these simulations it was possible to observe the efficiency of the algorithm in the reconfiguration of the RIS, managing to focus the electromagnetic waves in a remarkable way towards the point of interest.
\end{abstract}

\section{Introduction}
A metasurface is a planar destruction of metamaterial, a man-made material engineered to have electromagnetic properties not found in nature \cite{1, 2}. This material is made up of elementary units, arranged periodically in repeating patterns with dimensions and spacings much smaller than the wavelength \cite{4}. Thanks to these microscopic characteristics, chemical due to the nature of the resonant elements and geometric due to the dimensional nature, the metasurface is able to control the propagation of the electromagnetic waves with which it interacts, manipulating the distribution of the fields in space. Their ability to control the response of incident waves lies in the interaction of the electric and magnetic fields of the waves with the periodic structure of the surface which leads to resonances capable of changing the oscillatory nature of the electromagnetic fields \cite{7}. These interactions lead to resonant effects that lead to a wide range of applications such as broadband focusing \cite{9}, peculiar reflection/refraction \cite{10}, orbital angular momentum creation \cite{11}, space–surface wave manipulations \cite{12} and other applications in both microwave and optical frequencies. Although the metasurface has the ability to focus electromagnetic waves at various points in space, it remains a static structure, allowing it to always respond to the incident wave in the same way. Nowadays, therefore, the greatest technological interest has turned towards dynamic structures of metasurfaces capable of dynamically controlling the electromagnetic response of the surface. These surfaces are called Reconfigurable Intelligent Surfaces (RISs). The RIS is a type of surface that can change its configuration to affect signal propagation with low energy consumption managed by optimization software for the dynamic control of the propagation of electromagnetic waves \cite{13}. This can be done through the use of antenna elements which can be switched or adjusted to change the signal reflection and transmission behaviour \cite{14}. This allows to improve the performance of the communication system and to obtain greater flexibility in the design of the networks \cite{15}. The control of electromagnetic waves takes place through an intelligent reconfiguration of the wireless propagation environment which allows to improve the capacity and coverage of wireless networks \cite{16}. Their goal is to overcome the destructive effect of multi-path fading, which attenuates the communication signal strength, and improve the signal quality. Therefore, the RISs appear to be an advanced form of smart surface for the emerging hardware technology for the sixth-generation (6G) of radio communication networks \cite{17}. For RIS reconfiguration there are several optimization algorithms such as Several integrated software for metasurface reconfiguration such as Particle Swarm Optimization (PSO) \cite{18}, Network Slicing Resource Optimization (NSRO) \cite{19}, Salp Swarm Algorithm (SSA) \cite{20}, Gradient-based Optimization Algorithms which can be used to maximize signal reception in space. In this paper we analyze an open source multidimensional minimization algorithm from GNU Scientific Library (GSL) as the multivariable optimization algorithm for the RIS reconfiguration for smart radio environment.

\section{Simulation Set-Up}
\begin{figure}[t]
  \centering
  \includegraphics[width=82mm]{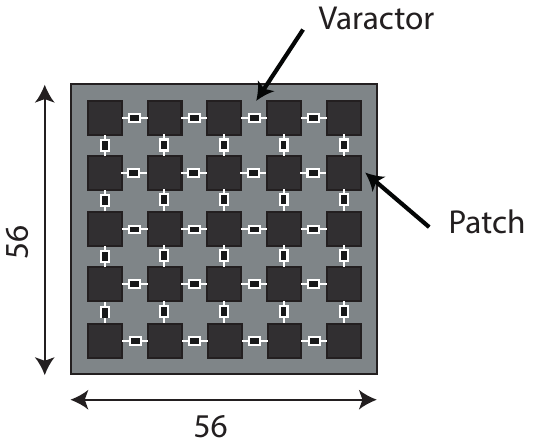}
  \caption{Representation of the simulated 56 $\times$ 56 mm $^2$ RIS. It consists of 5 $\times$ 5 PEC patches of 1 cm $^2$ connected by 40 varactor diodes, vertically and horizontally polarized.The dielectric support of $\epsilon_r=3$ and 1 mm of thickness.}
  \label{fig:fig1}
\end{figure}

The electromagnetic simulations are based on the analysis of the RIS focusing performance by its dynamic re-configurations in electromagnetic simulations. In order to analyze the EM response of the RIS reconfiguration, numerical simulations were run according to the finite difference time domain (FDTD) method to reproduce computational electrodynamics. The entire FDTD code has been implemented following the standard formulation written in C-language code. The electromagnetic simulations were performed in a domain of 140x140x200 cells of 1 mm cell size. The domain includes the RIS, two antennas, one transmitting antenna (Tx) and one receiving antenna (Rx) separated by a perfect electric conductor (PEC) metallic barrier. The antennas are dipole antennas of 10 cells working at  frequency band 2500-5000 MHz. The barrier, instead, consists of 1x60x60 cells placed in the middle of the two antennas in order to avoid the direct communication between them. The main characteristic of the barrier is the PEC boundary condition that consists of zero tangent electric field, that does not allow to transmit signal through the barrier. In this way, the only contribution that will allow the signal from the Tx to reach the Rx antenna will reside in the RIS. The simulated RIS is a periodic structure made up of 5x5 perfectly conductive resonant units (PECs) connected to each other by means of varactor diodes. The structure has dimensions equal to $56\times 1 \times 56$ cells, in which the patches are separated by 1 cell. Each patch communicates with two patches via variable capacitance varactor diodes for a total of 40 diodes in a dielectric film of 1 cell with $\epsilon_r=3$, separated from margins of 1 cell. This RIS is positioned in front of the dividing barrier, representing the only contribution for the receiving antenna. The entire configuration is shown in Figure \ref{fig:fig1}. The dielectric material of the interior of the chamber is air while the walls of the simulation space are perfect matched layers (PML).

\begin{figure}[htbp]
  \centering
  \includegraphics[width=69mm]{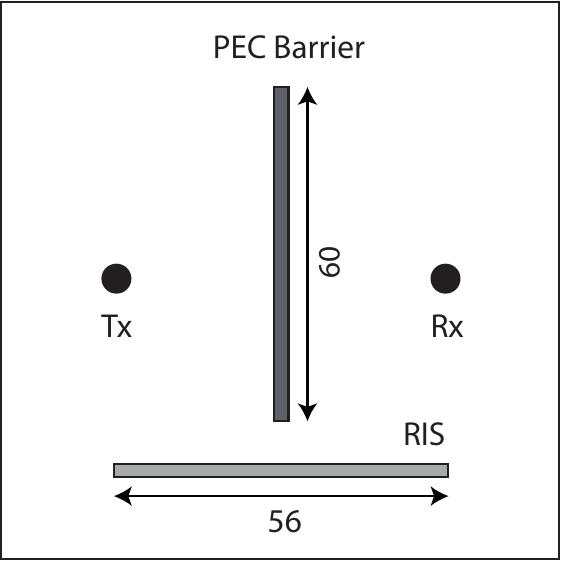}
  \caption{Representation of the simulation configuration: the RIS is positioned in the upper center of the workspace, the PEC barrier in the lower center while the two Tx and Rx antennas are located on the sides of the barrier. All the measures are in mm.}
  \label{fig:fig2}
\end{figure}

\section{Multidimensional Minimization}
The GNU Scientific Library is a set of open source numerical routines for scientific computing written in C and C++. It includes different kind of optimization algorithms that can be implemented in order to find the minimum or maximum of a given function. Among the optimization algorithm available in GSL, the multidimensional minimization has been chosen. This choice lies in the number of vactor diodes each of which is characterized by its own capacity. The goal of the algorithm is to iteratively find the ideal configuration of the varactor capacities in order to have the maximum towards the receiving antenna. The optimization function called multidimensional minimization looks for a point where the function to be optimized $f(x_1,..x_n)$ assumes the lowest value with respect to the neighboring points. To find the maximum, however, the objective is reversed to find ${1-\text{min}(f(x_1,..x_n))}$. The algorithm begins with a guess and descends using a search technique. A one-dimensional line is minimized using the gradient of the function until the lowest point is located within a reasonable tolerance. The process is repeated until the true n-dimensional minimum is found, at which point the search direction is updated using local information from the function and its derivatives. Until the simplex is small enough, the iteration process is continued. The process entails three steps: setting up the minimizer state s for algorithm T, updating s with iteration T, determining whether s has reached convergence, and repeating the iteration if necessary. Initializing the minimizer s to minimize the function fdf, the pre-built function initializes the multidimensional minimizer by starting from the initial point x. The tol parameter controls the accuracy of the line minimization, while the step-size parameter controls the size of the first trial step. If the gradient of the function g is orthogonal to the current search direction p and has a relative accuracy of tol, line minimization is frequently regarded as successful, with:
\begin{equation}
    p \cdot g < tol|p||g|
\end{equation}

\begin{figure*}[t]
  \centering
  \includegraphics[width=160mm]{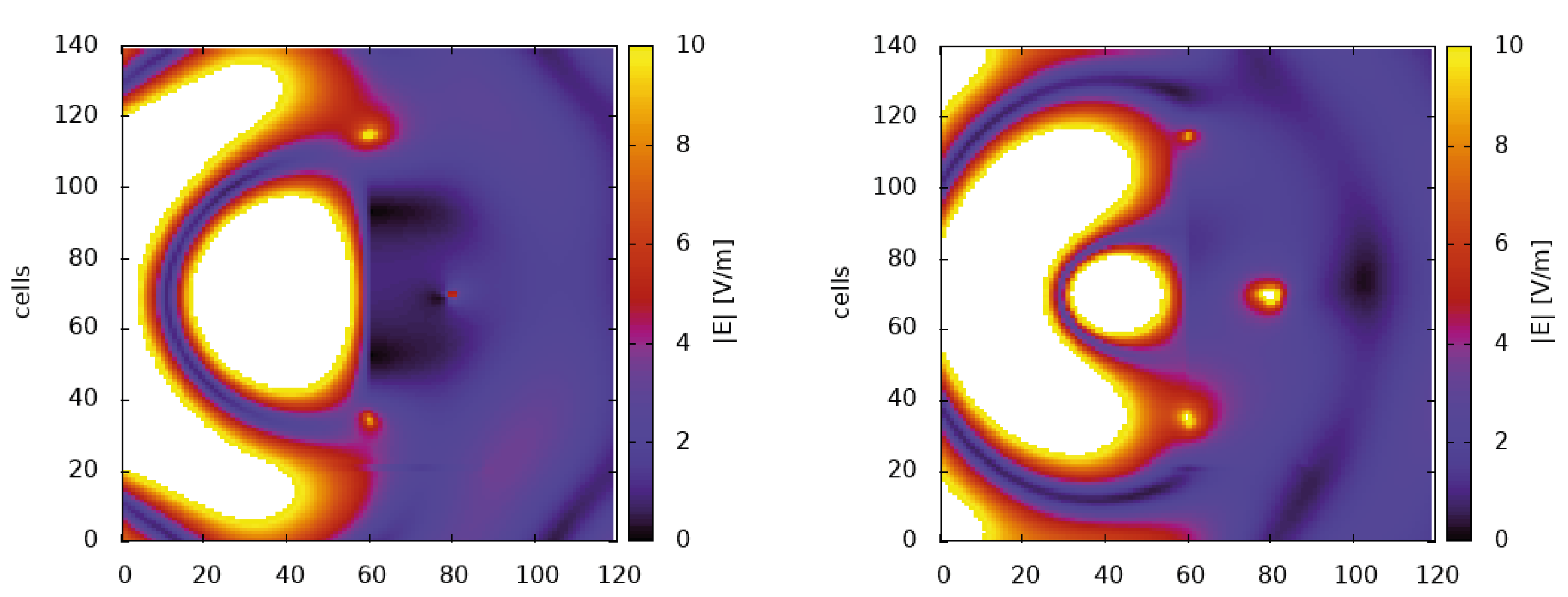}
  \caption{Electromagnetic field distribution in case of not reconfigured RIS (on the left) and in case of configured RIS (on the right).}
  \label{fig:fig3}
\end{figure*}

A tolerance value of 0.1 is chosen for the majority of tasks because the multidimensional minimization algorithm's line reduction process only needs to be performed roughly. A parametric function with n variables $f(x, \text{params})$ must be defined for the minimizers, along with two additional routines: one for calculating the function's gradient and the second for calculating both the function value and the gradient. The iteration functions are used to update the minimizer's state, and because the same function is used by all minimizers, different techniques can be swapped at runtime without modifying the code. The minimization process should stop when one of the following conditions is met: a minimum has been found within the user-specified precision, a user-specified maximum number of iterations has been reached, or an error has occurred. The minimization process compares the gradient g norm to the absolute tolerance epsabs. At a minimum, the gradient of a multidimensional function is zero. The test returns success if this condition is met:
\begin{equation}
    |g|<epsabs
\end{equation}
At this point the algorithm stops and the optimum set of parameters are then run on the last FDTD simulation for the best RIS reconfiguration.

\section{Results}
This section shows the results of the electromagnetic simulations according to the FDTD method considering the set-up shown in Figure \ref{fig:fig1}. The reported results refer to the electromagnetic response of the RIS before and after the metasurface reconfiguration. On the left in Figure \ref{fig:fig2}, it is shown the electromagnetic field distribution in case of not re-configured RIS. In this case the diodes are all switched on at fixed 1 pF capacitance, and how it is possibile to see there is not focusing toward the receiving antenna. On the right in Figure \ref{fig:fig2}, however, it is shown the electromagnetic field distribution in case of re-configured RIS.In this case, instead, there is greater focus near the reception point. 
\begin{figure}[htbp]
  \centering
  \includegraphics[width=82mm]{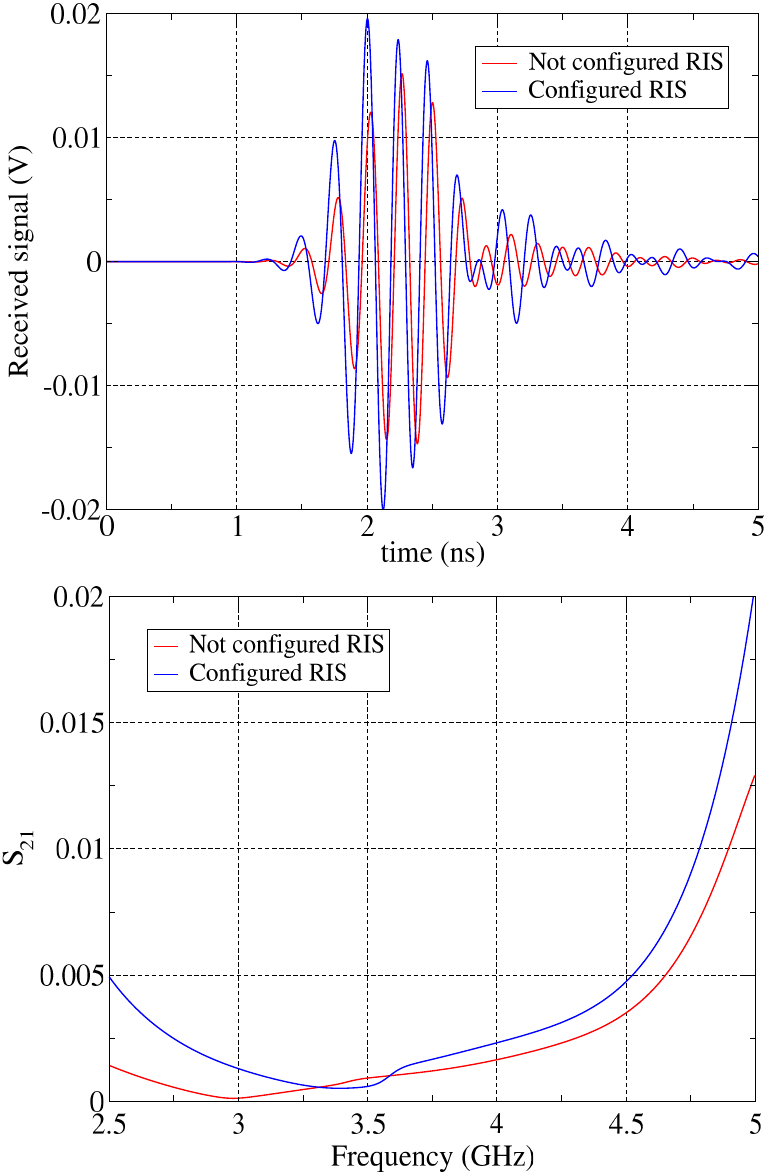}
  \caption{Received signal amplitude in time domain in case of not configured RIS (on top - red line) and configured RIS (top - blue line). The maximum value of the not configured RIS is 0.015 V while the configured one is 0.02 V. On the bottom is reported the $s_{21}$ in frequency domain in case of not configured RIS (red line) and configured RIS (blue line) from 2.5 GHz up to 5.0 GHz.}
  \label{fig:fig4}
\end{figure}
This distribution of fields focused towards the receiving antenna significantly improves the received signal. To quantify the signal received in the two configurations it is possible to observe the Figure \ref{fig:fig3}. On the left, the received signal in both reconfiguration is reported. The signal amplitude in time domain is shown on the left. As can be seen, in case of not reconfigured RIS (red line) the received signal amplitude is lower than in case of reconfigured RIS (blue line). On the right, instead, is reported the $s_{21}$ in frequency domain, in order to analyze the spectrum of the received signal. From 2.5 GHz up to 3.25 GHz, the received signal of the reconfigured RIS is higher as from 3.25 GHz up to 5 GHz with respect to the not reconfigured RIS. 

\section*{Acknowledgements}

This work has been supported by EU H2020 RISE-6G
project under the grant number 101017011.

\section{Conclusion}
In conclusion, the GNU Scientific Library's multidimensional optimization algorithm is effective at finding the optimal configuration for a specific point of focus. The algorithm has also two variations, the Fletcher-Reeves conjugate gradient algorithm and the Polak-Ribiere conjugate gradient algorithm. Both methods use a sequence of search directions to approximate the curvature of the function near the minimum. The Fletcher-Reeves method uses one approach for the coefficient beta, while the Polak-Ribiere method uses a different approach. Next steps foresee the analysis of the two versions of the multidimensional algorithm to guarantee a better reconfiguration of the metasurface. Upcoming activities are focused on analyzing different optimization algorithms and setups for the best reconfiguration of the RIS in smart radio environments.

\end{document}